\documentclass[aps,prb,twocolumn,superscriptaddress, longbibliography]{revtex4-2}

\pdfoutput=1

\usepackage{graphicx}
\usepackage{amsmath,amsfonts,amssymb}
\usepackage{color}
\usepackage{subfigure}
\usepackage{physics}
\usepackage{dsfont}
\usepackage{hyperref}
\usepackage{url}
\usepackage{leftidx}
\usepackage{textcomp}
\usepackage{gensymb}
\usepackage[nolist,nohyperlinks]{acronym}
\usepackage[normalem]{ulem}

\definecolor{emerald}{rgb}{0.31, 0.78, 0.47}
\definecolor{blue(ncs)}{rgb}{0.0, 0.53, 0.74}
\definecolor{emerald2}{rgb}{0.25, 0.73, 0.57}
\definecolor{benjoscolour}{rgb}{1.00, 0.50, 0.00}
\definecolor{orchid}{rgb}{0.6, 0.2, 0.8}

\hypersetup{
     colorlinks=true,
     linkcolor=blue,
     filecolor=blue,
     citecolor=emerald,      
     urlcolor =blue(ncs),
}

\DeclareMathAlphabet{\pazocal}{OMS}{zplm}{m}{n}

\newcommand{\bo}[1]{\boldsymbol{#1}}
\newcommand{\D}{\mathrm{d}}
\begin{document}

\title{Axion-matter coupling in multiferroics}

\author{Henrik S. R{\o}ising}
\email{henrik.roising@su.se}
  \affiliation{Nordita, KTH Royal Institute of Technology and Stockholm University, Hannes Alfv\'{e}ns v\"{a}g 12, SE-106 91 Stockholm, Sweden}
  
\author{Benjo Fraser}
  \affiliation{Nordita, KTH Royal Institute of Technology and Stockholm University, Hannes Alfv\'{e}ns v\"{a}g 12, SE-106 91 Stockholm, Sweden}
  
 \author{Sin\'{e}ad M. Griffin}
   \affiliation{Materials Sciences Division, Lawrence Berkeley National Laboratory, Berkeley, California 94720, USA} 
   \affiliation{Molecular Foundry, Lawrence Berkeley National Laboratory, Berkeley, CA 94720, USA}
   
\author{Sumanta Bandyopadhyay}
  \affiliation{Nordita, KTH Royal Institute of Technology and Stockholm University, Hannes Alfv\'{e}ns v\"{a}g 12, SE-106 91 Stockholm, Sweden}

\author{Aditi Mahabir}
 \affiliation{Department of Physics, University of Connecticut, Storrs, Connecticut 06269, USA} 
 
 \author{Sang-Wook Cheong}
 \affiliation{Rutgers Center for Emergent Materials and Department of Physics and Astronomy, Rutgers University, Piscataway, New Jersey 08854, USA}
   
\author{Alexander V. Balatsky}
\email{balatsky@hotmail.com}
  \affiliation{Nordita, KTH Royal Institute of Technology and Stockholm University, Hannes Alfv\'{e}ns v\"{a}g 12, SE-106 91 Stockholm, Sweden}
 \affiliation{Department of Physics, University of Connecticut, Storrs, CT 06269, USA} 
  
\date{\today}

\begin{abstract}
Multiferroics (MFs) are materials with two or more ferroic orders, like spontaneous ferroelectric and ferromagnetic polarizations. Such materials can  exhibit a magnetoelectric effect whereby magnetic and ferroelectric polarizations couple linearly, reminiscent of, but not identical to the electromagnetic $\boldsymbol{E}\cdot \boldsymbol{B}$ axion coupling. Here we point out a possible mechanism in which an external dark matter axion field couples linearly to ferroic orders in these materials without external applied fields. We find the magnetic response to be linear in the axion-electron coupling. At temperatures close to the ferromagnetic transition fluctuations can lead to an enhancement of the axion-induced magnetic response. Relevant material candidates such as the Lu-Sc hexaferrite family are discussed.
\end{abstract}

\maketitle

%
%%
%%%
\section{Introduction}
\label{sec:Intro}
%%%
%%
%

Dark matter (DM) is believed to comprise over three quarters of the universe's mass-density, yet so far it eludes detection in spite of the intensive search using numerous detection schemes~\cite{Feng10}. One possible explanation is that the DM particles are too light to strongly scatter off nuclei and electron-based detection schemes currently proposed, motivating the need for new ideas to explore the sub-GeV range of DM masses. In this regard, the  $\mathcal{O}$(meV) axion, a pseudoscalar boson introduced to solve the strong charge-parity problem in quantum chromodynamics~\cite{PecceiQuinn77, Weinberg78, Wilczek78, KimEA10}, offers a particularly well-motivated DM candidate~\cite{PreskillEA83, AbbottSikivie83, DineFischler83, DuffyEA09, Ringwald12}. 

A new avenue to DM detection is offered by quantum materials~\footnote{We adopt the convention of referring to materials which essential properties are inherently described by quantum mechanics, such as strongly correlated electron systems, as \emph{quantum materials}.} in which the energy scales of the excitations coincide with the requirements for lower-mass DM detection. Moreover, quantum materials possess entangled collective degrees of freedom that allow for an enhanced coupling to axions. Traditionally, the majority of axion detection proposals have focused on exploiting the axion-photon coupling in cavities or plasma haloscopes using strong external fields~\cite{Sikivie83, IrastorzaEA18, PDG18, LawsonEA19, AlesiniEA17, Sikivie20, Rybka14, BrusbakerEA17, BudkerEA14, CaldwellEA17, GoryachevEA18}.  More recently, axion physics has been increasingly discussed in the condensed matter context~\cite{NennoEA20, Gramolin20, Frey20, Gramolin20, MitridateEA20} primarily in analogy to axion electrodynamics in topological insulators~\cite{EssinEA09, LiEA10, WangEA16, TaguchiEA18, Armitage19}, but also in terms of particle axion detection via resonant coupling to quasiparticles~\cite{MarshEA19, EngelEA21}. In most of the existing proposals, the detection scheme uses DM particle absorption to induce single particle excitations in the sensor material, e.g., the particle-hole excitations in a small gap material.  
\begin{figure}[t!bh]
	\centering
	\includegraphics[width=0.85\linewidth]{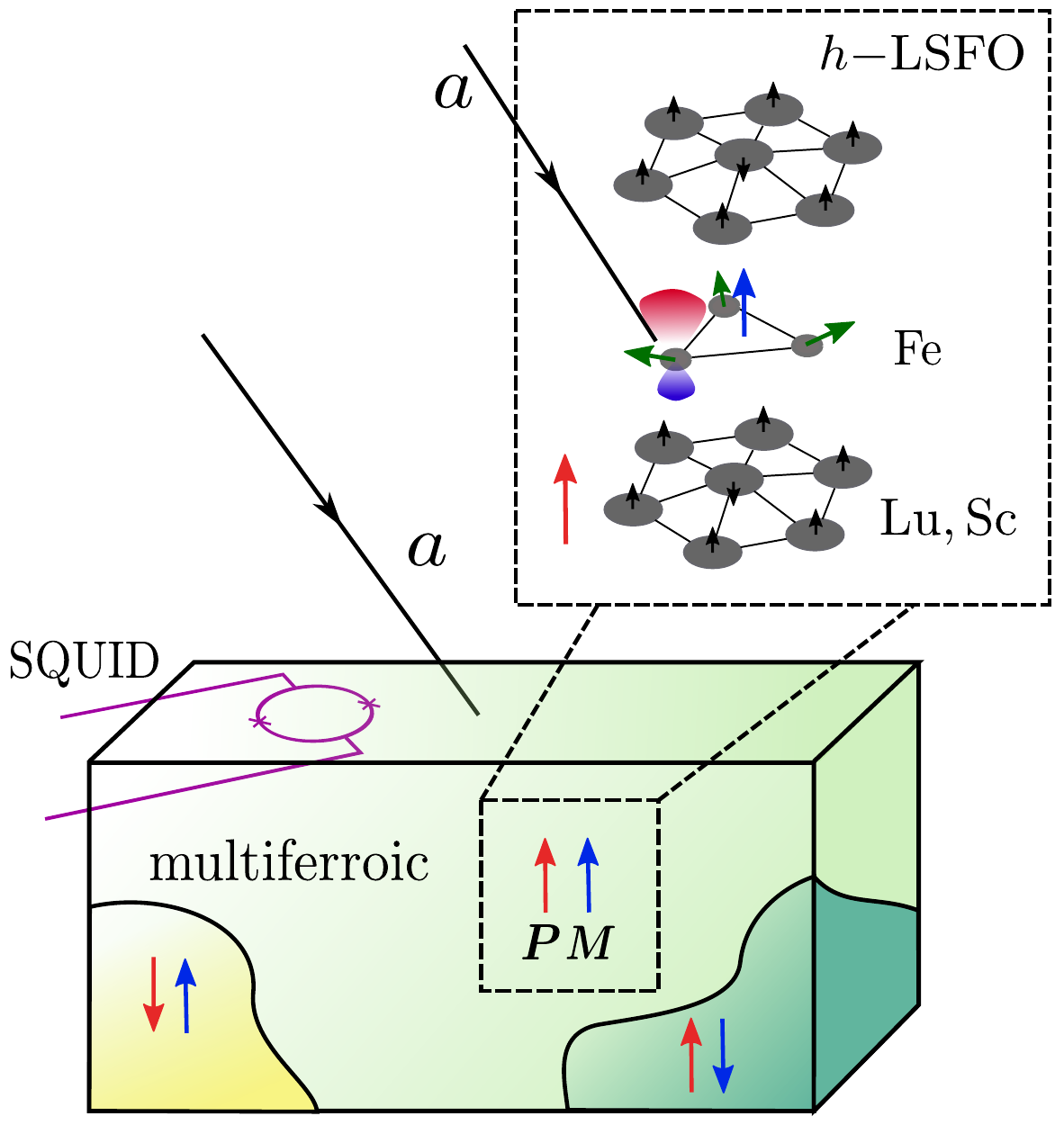} 
	\caption{An incident axion $a$ couples to multiferroic orders via the electron spin and momentum below the Curie temperature. Close to the onset of ferromagnetism, softness in the magnetization makes the system susceptible to the axion coupling. The magnetic material response is imagined sensed by a sensitive magnetometer such as a superconducting quantum interference device (SQUID). The inset shows a sketch of the $A_2$ phase of multiferroic Lu$_{1-x}$Sc$_x$FeO$_3$ ($h$-LSFO)~\cite{DuEA18}.}
	\label{fig:Idea}
\end{figure}
An alternative approach is to detect DM  by their coupling to collective modes of the quantum material. With the rapid progression of the field of axion electrodynamics, we anticipate more such proposals for axion sensing schemes will be explored.

Here we explore multiferroics (MFs) as a platform for DM detection using materials with parallel ferroic orders $\bo{P} \parallel \bo{M}$ as an ``impedance matched" medium to interact with axion DM. Here, $\bo{P}$ ($\bo{M}$) is the electric (magnetic) polarization vector. Multiferroics with $\bo{P} \parallel \bo{M}$ act as the ``condensate'' axion field seen by an arriving external axion. The external axion field linearly couples to the polarization fields and both changes net magnetization and excites the collective modes of the MF. The mechanism for DM detection proposed here is qualitatively different from existing schemes in the sense that the axion DM affects the magnetization---a macroscopic observable---by modifying the parameters that control the macroscopic state. As such, the proposed method of detection is similar to the transition-edge sensors, except that our sensor (the multiferroic sample) remains in a macroscopically ordered state before and after interaction with DM. 

To place the current discussion in a broader context: One finds two main developing threads for axion DM sensors. One approach is relying on the sensor device being placed in external electric ($\bo{E}$), magnetic ($\bo{B}$) or both fields simultaneously, whereby one breaks time reversal ($\pazocal{T}$) and parity ($\pazocal{P}$) to induce coupling to the axion field. Another approach is to use systems that take advantage of spontaneous symmetry breaking (no external fields applied). In the phase with broken $\pazocal{P}$ and $\pazocal{T}$ symmetry coupling of the axion DM field to {\em matter fields} occurs naturally. The coupling induces collective excitations that are subsequently read out by appropriate sensing means. Here we propose a new realization of the latter platform where $\pazocal{P}$ and $\pazocal{T}$ breaking (though with $\pazocal{P}\pazocal{T}$ being a symmetry operation), and hence axion electrodynamics, occurs naturally. 

Our axion-matter sensing proposal is sketched in Fig.~\ref{fig:Idea}. Our central scheme is that in MFs with \emph{parallel} ferroic orders in zero external fields one has a medium that \emph{linearly} couples to an axion field $a$ through an $a \bo{P} \cdot \bo{M}$ term in the  Hamiltonian. This term allows for a novel axion DM detection scheme: Impinging axions will hybridize and directly couple with ferroic orders in MFs without requiring an intermediate photon coupling. This coupling between the axion field and matter fields (with $\bo{P} \parallel \bo{M}$) is a {\em bulk} effect. The electrons in the MF produce a $\pazocal{P}$ and $\pazocal{T}$ breaking term that can be expressed in terms of $\bo{P}$ and $\bo{M}$ but which microscopically results in the bulk expectation value of the axion-electron coupling as described in Sec.~\ref{sec:Matterfields}. The macroscopically ordered state provides a set of sensors where each ``atomic'' unit interacts with the same axion field, and their contributions add coherently---the signal scales with a macroscopic number of atoms. Phrased differently, we obtain \textit{Avogadro scaling} due to the $\bo{P} \parallel \bo{M}$ coherent state of the material.  The proposed scheme thus offers a quantum advantage at the macroscopic level. This ``Avogadro advantage'' is a unique feature of macroscopically coherent quantum matter where each unit cell, or each part of the coherent condensate, adds to increased detection sensitivity. We see this advantage operational for the numerous detection schemes relying on the macroscopically coherent states ranging from ferromagnets to superfluid/superconducting condensates and MFs. To discuss multiferroics as a proposed platform in this context we will use the acronym \acronym{MERMAID} (multiferroic matter axion detector).

We also present a known material possessing  our required coupling to give materials-specific estimates of our MERMAID scheme. We focus on the recently discovered hexagonal LuFeO$_{3}$ thin films and the Lu$_{1-x}$Sc$_x$FeO$_3$ crystal family, which have a ferromagnetic ground state for $x\approx 0.4$ below $T_{C_M} \approx 160$~K~\cite{LinEA16, DuEA18}. Sample candidates may be synthesized and characterized using THz magneto-optical spectroscopy to determine the nature of the collective excitations, including electromagnons, and the resulting $\bo{P} \parallel \bo{M}$ coupling strength.

This paper is organized as follows. We provide background on axions in Sec.~\ref{sec:Axions}. An effective theory of the axion-matter coupling in multiferroics is presented in Sec.~\ref{sec:EffectiveTheory}. In Sec.~\ref{sec:Matterfields} we discuss the effective axion-electron coupling in multiferroics, and we pursue an effective Ginzburg--Landau theory to describe the dynamical response in Sec.~\ref{sec:FreeEnergy}. Critical slowing down and the axion-induced magnetic response are discussed in Sec.~\ref{sec:CritSlowdown} and \ref{sec:Sensitivity}, respectively. In Sec.~\ref{sec:AScaling} we discuss the advantages of Avogadro scaling in multiferroics. In Sec.~\ref{sec:Materials} we discuss material candidates and estimates from \emph{ab initio} density functional theory calculations for a specific compound. We provide conclusions and an outlook in Sec.~\ref{sec:Conc}.
 
%
%%
%%%
\section{Background: Axions}
\label{sec:Axions}
%%%
%%
%
The axion is an hypothesized pseudoscalar boson originally proposed to solve the strong charge-parity problem via the Peccei--Quinn mechanism in quantum chromodynamics (QCD)~\cite{PecceiQuinn77, Weinberg78, Wilczek78, Wilczek87}. It was realized that the axion could be a viable DM candidate, which has significantly fuelled interest in it~\cite{PreskillEA83, AbbottSikivie83, DineFischler83, DuffyEA09}. The axion couples to electromagnetism (in units where $c = \hbar = 1$) via 
\begin{align}
\pazocal{L}_a &= \frac{1}{2} (\partial_{\mu} a)(\partial^{\mu} a) - \frac{1}{2} m_a^2 a^2, \label{eq:Lagrangian} \\
\pazocal{L}_{\gamma \gamma a} &= - \frac{\theta}{4} F_{\mu \nu} \tilde{F}^{\mu \nu} = \theta \bo{E} \cdot \bo{B}, \label{eq:EMCoupling} \\
\theta &\equiv a g_{\gamma \gamma a},  \label{eq:ThetaDef}
\end{align}
where $F^{\mu \nu}$ is the electromagnetic tensor with $\tilde{F}^{\mu \nu} = \frac{1}{2}\varepsilon^{\mu \nu \alpha \beta} F_{\alpha \beta}$ its dual ($\varepsilon^{\mu \nu \alpha \beta} $ is the Levi--Civita symbol), and where $a$ ($m_a$) is the axion field (mass), and $g_{\gamma \gamma a}$ is a dimensionful coupling strength.

Although axion-like particles with a Lagrangian like Eqs.~\eqref{eq:Lagrangian} and \eqref{eq:EMCoupling} within a vast mass range could account for a partial DM abundance, QCD axions have the particular coupling strength $g_{\gamma \gamma a} = \pazocal{C}_a \frac{m_a}{\text{GeV}^2}$, where $\pazocal{C}_a$ is a model dependent number that ranges between $-0.39$ and $0.15$ for various theories~\cite{diCortonaEA16}. Despite an in principle still vast $(m_a, g_{\gamma \gamma a})$ parameter space, the axion mass is restricted by $m_a \lesssim 0.1$~eV from astrophysical bounds~\cite{PDG18} and by $m_a \gtrsim 10^{-12}~$eV from cosmological bounds~\cite{PreskillEA83, AbbottSikivie83, DineFischler83}. Moreover, the coupling strength is roughly limited to $\lvert g_{\gamma \gamma a} \rvert < 10^{-10}~$GeV$^{-1}$ by helioscope experiments~\cite{PDG18}. Interest in the mass range around meVs has been fuelled by extensive lattice QCD computations~\cite{BorsanyiEA16} and searches at the XENON experiment~\cite{GaoEA20}.

Moreover the axion couples to Dirac fermions via~\footnote{A non-derivative fermion coupling $i\,a\,\bar{\psi} \gamma^5 \psi$ is often discussed in the literature. However by a suitable chiral rotation of the fermion field we can take \eqref{eq:FermionCoupling} to be the tree level interaction, without loss of generality.}
\begin{equation}
\pazocal{L}_{af} = - g_{af}  \frac{\partial_{\mu} a}{2m_f} \bar{\psi} \gamma^5 \gamma^{\mu} \psi,
\label{eq:FermionCoupling}
\end{equation}
where $\psi$ is the Dirac spinor and $g_{af}$ is a dimensionless coupling strength. The coupling to electrons is limited by $\lvert g_{ae} \rvert < 3\times 10^{-11}$ from solar neutrino experiments~\cite{GondoloEA09, BarthEA13}. The relevant low-energy Hamiltonian resulting from Eq.~\eqref{eq:FermionCoupling} is derived in Appendix~\ref{sec:AxionFermion}. We note that this Hamiltonian contains an effective term of the form $\theta \bo{P} \cdot \bo{M}$, which is present even in the absence of external fields ($\bo{A} = 0$), see Eq.~\eqref{eq:HamEffective}. 

Since the axion is expected to be light, we can treat $a$ as a classical field with a long de Broglie wavelength compared to the typical experimental probe length scale to give an estimate of $\theta$ in Eq.~\eqref{eq:ThetaDef} for QCD axions. We assume that the the axion field is oscillating with a frequency set by its mass, $a = a_0 \exp(i m_a t)$. Assuming that the axion makes up the local DM density~\cite{MillarEA17, CatenaUllio10, PDG18}, $\rho_{\text{DM}} \approx 300~\text{MeV} \text{cm}^{-3}$, the amplitude $a_0$ is fixed by $\rho_{\text{DM}} = \frac{1}{2} m_a^2 \lvert a_0 \rvert^2$, yielding e.g.,~$\lvert \theta \rvert \approx \lvert a_0 g_{\gamma \gamma a} \rvert \sim 10^{-22}$.

%
%%
%%%
\section{Effective Theory}
\label{sec:EffectiveTheory}
%%%
%%
%

In this section we develop a framework to study the magnetic response in MFs mediated by an axion-fermion coupling. In Sec.~\ref{sec:Matterfields} we discuss how an effective matter analogue of the axion $\bo{E}\cdot \bo{B}$ coupling emerges in MF systems, with details of the derivation listed in Appendices~\ref{sec:AxionFermion}, \ref{sec:AxionFermionFEv2}, and \ref{sec:rel}.

\subsection{Multiferroics and effective matter coupling} 
\label{sec:Matterfields}

With a matter analogue of Eq.~\eqref{eq:EMCoupling} in mind, multiferroics are materials in which spontaneous electric and magnetic polarizations can develop~\cite{WangEA09, TokuraEA14, FiebigEA16, DuEA18, SpaldinRmesh19, ChengliangEA19}. In solid-state crystals, ferroelectricity is normally driven by lone-pair, geometric, charge ordering, or spin-driven mechanisms~\cite{FiebigEA16}, which result in a macroscopic breaking of inversion symmetry. As a result of spontaneous symmetry breaking in the MF bulk of both time reversal and inversion symmetry, there exist internal electric and magnetic fields (even in the absence of external fields) resulting from these symmetry-broken order parameters. Based on symmetry considerations one should expect an effective $\chi \theta \bo{P}\cdot \bo{M}$ coupling to be present in the MF phase, despite this term being commonly overlooked in the high-energy literature. 

Microscopically, an effective $\bo{P} \cdot \bo{M}$ axion term can be derived from the coupling to electrons, as per Eq.~\eqref{eq:FermionCoupling}, in the MF ground state. For details we refer to Appendices~\ref{sec:AxionFermion} and~\ref{sec:AxionFermionFEv2}. The key result is that the axion-electron coupling at low energy contains the term
\begin{equation}
\delta E_{ae} \approx - i \hbar  g_{ae} \frac{\partial_t a}{2m_e^2 c} \bra{\Psi}\bo{\sigma} \cdot \bo{\nabla} \ket{\Psi},
\label{eq:AxionElectronLowE} 
\end{equation} 
where $\ket{\Psi}$ is the MF ground state. In the MF phase Eq.~\eqref{eq:AxionElectronLowE} is dictated by the distortions of electronic orbits (densities) such that one obtains a spontaneous magnetic and electric dipole moment throughout the whole sample.  In a (rare) subclass of MFs these moments are parallel, as discussed in Sec.~\ref{sec:Materials}. Despite the $m_a/m_e$ suppression, coming from the $\mu = 0$ contribution to Eq.~\eqref{eq:FermionCoupling}, we argue in Sec.~\ref{sec:AScaling} that the energy of Eq.~\eqref{eq:AxionElectronLowE} can add coherently over a macroscopic volume limited by the MF domain size, which can result in an appreciable effect. To be clear, the interaction in Eq.~\eqref{eq:AxionElectronLowE}, evaluated on electronic states, yields a {\em bulk} contribution that is  evaluated by model calculations (Sec.~\ref{sec:AScaling}) and \emph{ab initio} means (Sec.~\ref{sec:Materials}). In Appendix~\ref{sec:AxionFermion} we discuss the relation to other couplings that are considered in existing detection schemes~\cite{Sikivie14, Hill16, MitridateEA20}.

\subsection{Ginzburg--Landau theory}
\label{sec:FreeEnergy}
Using an isotropic sample bulk model, we proceed with an effective description of the longitudinal dynamics of the internal MF degrees of freedom. This approach is valid close to the transition temperature where the free energy can be expanded in powers of the respective order parameters. In terms of the free energy density $\pazocal{F}$, $F[\bo{M}, \bo{P}] = \int \D^3 r~\pazocal{F}[\bo{M}, \bo{P}]$, in the absence of external electric and magnetic fields, we have (in SI units)
\begin{widetext}
\begin{equation}
\begin{aligned}
\pazocal{F}[\bo{M}, \bo{P}] &= \mu_0 \pazocal{F}_M[\bo{M}] + \frac{1}{\varepsilon_0}  \pazocal{F}_P[\bo{P}] - \sqrt{\frac{\mu_0}{\varepsilon_0}}(\alpha_{ij} M^{i} P^{j} + \chi \theta  \bo{P} \cdot \bo{M} ) , \\
\pazocal{F}_M[\bo{M}] &= \alpha_M \lvert \partial_t \bo{M} \rvert^2 - \beta_M \lvert \bo{\nabla} \cdot \bo{M} \rvert^2 -  \gamma_M(T-T_{C_M}) \lvert \bo{M} \rvert^2 - \lambda_M \lvert \bo{M} \rvert^4 + \dots, \\
\pazocal{F}_P[\bo{P}] &= \alpha_P \lvert \partial_t \bo{P} \rvert^2 - \beta_P \lvert \bo{\nabla} \cdot \bo{P} \rvert^2 - \gamma_P(T-T_{C_P}) \lvert \bo{P} \rvert^2 - \lambda_P \lvert \bo{P} \rvert^4 + \dots,
\end{aligned}
\label{eq:FreeEnergy}
\end{equation}
\end{widetext}
where $\varepsilon_0$ ($\mu_0$) is the vacuum permittivity (permeability), and where $\alpha$'s, $\beta$'s, $\gamma$'s, and $\lambda$'s are phenomenological coefficients, $\alpha_{ij}$ is the material dependent ME tensor, and $\chi$ is a material dependent susceptibility. The transition temperature $T_{C_M}$ ($T_{C_P}$) denotes the magnetic (electric) Curie temperature above which the ferromagnetic (ferroelectric) order melts. In analogy to the $\pazocal{P}\pazocal{T}$-odd axion field, where $\pazocal{P}$ ($\pazocal{T}$) is the inversion (time reversal) symmetry operator, the ME tensor is $\pazocal{P}\pazocal{T}$-odd if the system (crystal) otherwise respects inversion and time reversal symmetry. The magnitude of the elements $\alpha_{ij}$ are limited by thermodynamic stability of the theory~\cite{SpaldinEA08}.

We consider the case $T\approx T_{C_M} \ll T_{C_P}$ (as applicable to the majority of MF materials). This makes $\bo{P}$ act as a ``hard'' and $\bo{M}$ as a ``soft'' field; for the latter fluctuations are greater. Hence we take the electric polarization to be close to its minimum energy value, and consider fluctuations, $\gamma_P(T-T_{C_P}) \lvert \bo{P} \rvert^2 + \lambda_P \lvert \bo{P} \rvert^4 \to \frac12 m_P^2 \lvert \delta \bo{P} \rvert^2$, where $\delta \bo{P}$ are the ferroelectric fluctuations, $\bo{P} = \bo{P}_0 + \delta \bo{P}$, and where $m_P$ is the effective mass. Below we relabel $\delta \bo{P} \to \bo{P}$ and bear in mind that $\bo{P}$ is a fluctuation field. When associating $\pazocal{F}$ in Eq.~\eqref{eq:FreeEnergy} with a Lagrangian density it yields the classical equations of motion for $\bo{M}$ and $\bo{P}$:
\begin{align}
\alpha_M \partial_t^2 \bo{M} - \beta_M \nabla^2 \bo{M} + \gamma_M (T-T_{C_M}) \bo{M} &  \nonumber \\
+ \frac{c}{2}(\underline{\alpha} + \chi \theta) (\bo{P}+\bo{P}_0) + 2\lambda_M \lvert \bo{M} \rvert^2 \bo{M} &= 0, \label{eq:ELeqM} \\
\alpha_P \partial_t^2 \bo{P} - \beta_P \nabla^2 \bo{P} + \frac12 m_P^2\bo{P} + \frac{1}{2c}(\underline{\alpha} + \chi \theta) \bo{M} &= 0, \label{eq:ELeqP}
\end{align}
where $c$ is the speed of light and $\underline{\alpha}$ is the ME tensor with elements $\alpha_{ij}$. Turning off both $\underline{\alpha}$ and $\theta$ reduces Eqs.~\eqref{eq:ELeqM} and \eqref{eq:ELeqP} to two uncoupled Klein--Gordon equations when ignoring the cubic (Gross--Pitaevskii) term, which is justified when $T > T_{C_M}$.

We emphasize that the above effective theory for the magnetic order is a classical and minimal description of the longitudinal magnon dynamics. In situations where the transverse modes are involved, dynamics from the Landau--Lifshitz--Gilbert equation is expected to play a role~\cite{Leutwyler94}. Finally, we note that in a crystal the Ginzburg--Landau theory away from low filling should respect symmetries of the crystal point group and partition the solution into irreducible representations. This layer of complexity is omitted in the isotropic bulk theory considered here and would constitute a natural extension of the theory. 

\subsection{Critical slowing down}
\label{sec:CritSlowdown}
On general grounds we expect to have two regimes of the coupling following the above equations. The axion field provides a periodic perturbation set by the time scale $\tau_{a} \sim 1/\omega_a$, while the characteristic time of magnetization is set by the mass term $\tau \sim \lvert T - T_{C_M} \rvert^{-1/2}$, and more generally, the magnetization dynamics experiences a critical slowing down near $T_{C_M}$: 
\begin{equation}
 \tau \sim \lvert T - T_{C_M}\rvert^{-\nu z},
 \label{eq:CritSlowdown}
\end{equation}
\begin{figure}[t!bh]
	\centering
	\includegraphics[width=0.75\linewidth]{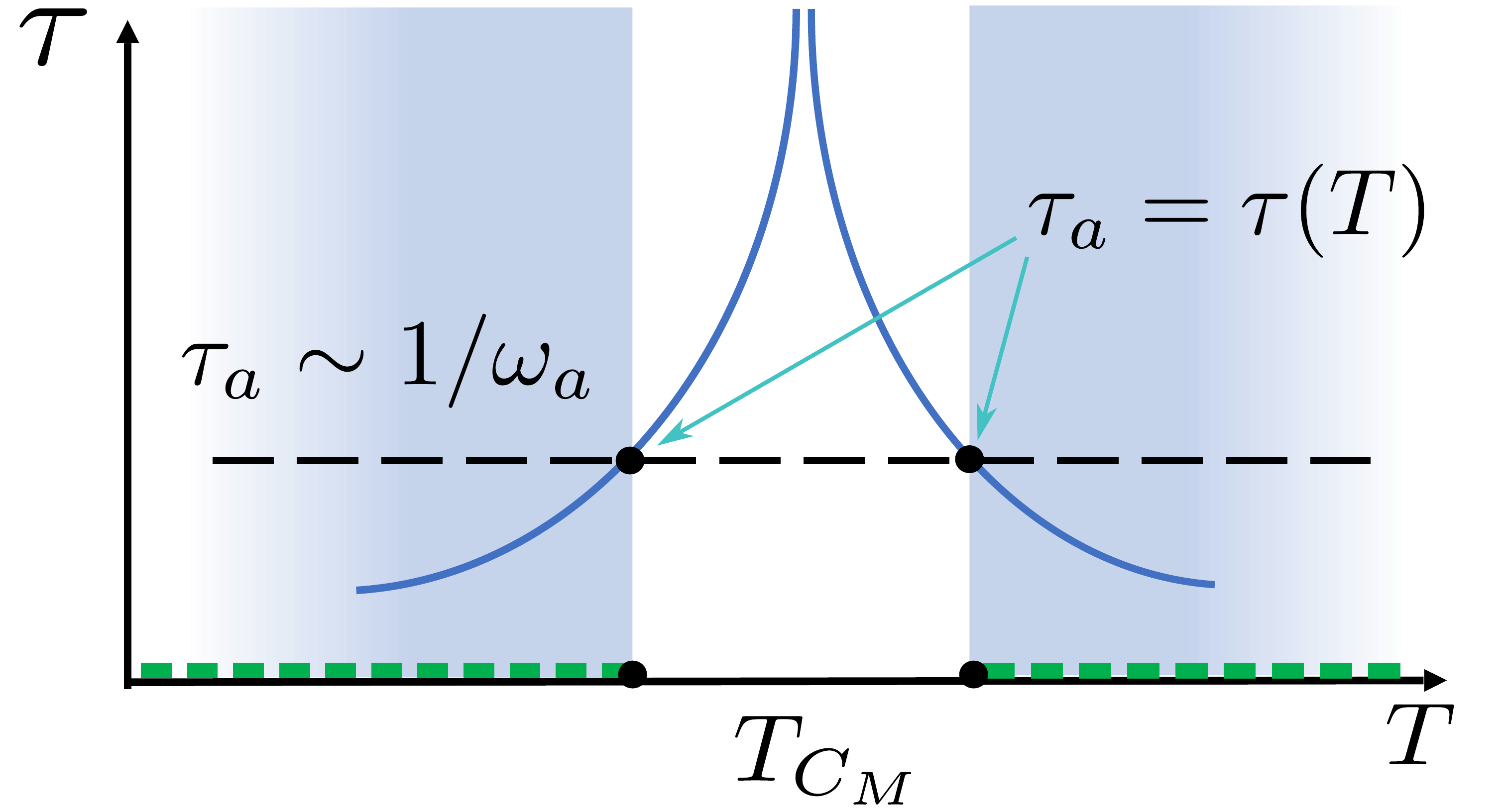}
	\caption{Sketch of critical slowing down for the magnetization dynamics time scale $\tau$ (blue curve). The horizontal dashed line indicates the steady axion dynamics time scale $\tau_a$. Temperature tuning can be used to match the magnetic time scale with the axion time scale. The adiabatic regime away from the critical point, where axion dynamics is slow compared to magnetization time scale $\tau < \tau_a$, is shown with a green dashed lines on the temperature axis.}
	\label{fig:CritSlowdown}
\end{figure}
for some dynamical critical exponent $z\nu \sim 1 $ determined by the nature of the transition~\cite{HolmEA93, HohenbergHalperin77, Cardy1996}. The (i) adiabatic regime of the axion-magnetization dynamics occurs at $\tau_{a} > \tau$. In this case the effective axion field seen by magnetization has a similar effect to a steady external field, and the response accumulates in time. In the opposite case of (ii) antiadiabatic evolution, $\tau_{a} < \tau$, the magnetization ``sees'' a rapidly oscillating $\theta(t)$ term that effectively averages to zero. The most direct and strongest effects of the axion field will be seen in the temperature regime where the axion field dynamics is slow compared to the magnetization time, i.e.,~regime (i). This requirement places a constraint on how close one can approach criticality, as illustrated in Fig.~\ref{fig:CritSlowdown}.

To give a numerical estimate of this constraint we assume that $\tau = \tau_0 (1-T/T_c)^{-\nu z}$ for $T < T_c$, where $\tau_0$ is material dependent. The crossover between the adiabatic and the antiadiabatic regime occurs at $T/T_c = 1 - (\tau_0 \omega_a)^{1/(\nu z)}$, given that $\tau_0\omega_a < 1$. For $m_a = 1~$meV this would require $\tau_0 \lesssim 0.5$~ps. Experiments conducted on ultrathin Fe films and multiferroic MnWO$_4$ have found $\tau_0$ in the range of $\tau_0 \sim 1~$ns~\cite{DunlavyEA05, NiermannEA15}, which if applied to our context would limit the crossover to occur for light axions $m_a \lesssim 1~\mu$eV. For instance, for $m_a = 0.1~\mu$eV and $\nu z = 1$ we find $T/T_c \approx 0.85$. For these estimates, which are merely indicative, one would thus want to maintain the system close to the temperature $T = 0.85~T_c$. In subsequent work we plan to undertake a more systematic analysis of relaxation times.

\subsection{Magnetic response}
\label{sec:Sensitivity}
The coupled dynamical equations \eqref{eq:ELeqM} and \eqref{eq:ELeqP} can be treated with linear response theory with $\theta = \theta_0 \exp(i\omega_a t)$ acting as a weak time-dependent driving force. Here we consider the result of such a treatment in one spatial dimension with $\alpha_{ij} = \alpha \delta_{ij}$ assumed for simplicity, with the details listed in Appendix~\ref{sec:Mathieu}. In short: We find that the axion acts as a periodic driving term to the dynamic material magnetization. The basic experimental signature of the axion would thus be an oscillatory magnetization of frequency $\omega_a$ on top of the static (ferromagnetic) background.

When $T < T_{C_M} \ll T_{C_P}$ and the magnetization is soft compared to the electric polarization, the classical magnetic response resembles that of a driven harmonic oscillator. In frequency space the magnetization has a linear response function of the form
\begin{equation}
    \Big\lvert \frac{\delta M}{\theta_0}\Big \rvert \, =\, \frac{1}{\sqrt{ \left(\omega_a^2 - \omega_M^2\right)^2+ \eta_M^2 \omega_a^2 }}\, \frac{\chi}{2\alpha_M} \Big\lvert \frac{\alpha M_0}{2 m_P^2} - c P_0 \Big\rvert,
    \label{eq:MagresponseMain}
\end{equation}
with $M = M_0 + \delta M$, $P \approx P_0$, where $M_0$ and $P_0$ are the static polarizations, and $\delta M$ are magnetic fluctuations. $\omega_M$ is the magnetic (magnon) frequency, and $\eta_M$ is a damping coefficient of a term $\eta_M \partial_t M$ added to simulate energy dissipation in the system. Addressing how low $\eta_M$ can be pushed and whether it is a limiting factor experimentally is outside the scope of this paper. However, for spin waves to be resolved we require $\eta_M \lesssim 0.01 J$, where $J \sim \pazocal{O}(1~\text{meV})$ is the typical spin-exchange interaction.

The expression Eq.~\eqref{eq:MagresponseMain} sheds light on the key factors in the response. First, there is the overall coefficient, i.e.~the factor inside the absolute value. This shows that there are two contributions to the driving: the parallel magnetoelectric effect $\alpha$, and the static electric polarization $P_0$. Since the $P_0$ term can reasonably be expected to dominate in many cases~\footnote{In the case of $h$-LSFO $c P_0 \approx 26~$kOe, so the ferroelectric contribution dominates the right-hand side of Eq.~\eqref{eq:MagresponseMain}.}, the ferroelectricity will be a controlling factor in the driving. 

\begin{figure}[t!bh]
	\centering
	\includegraphics[width=0.95\linewidth]{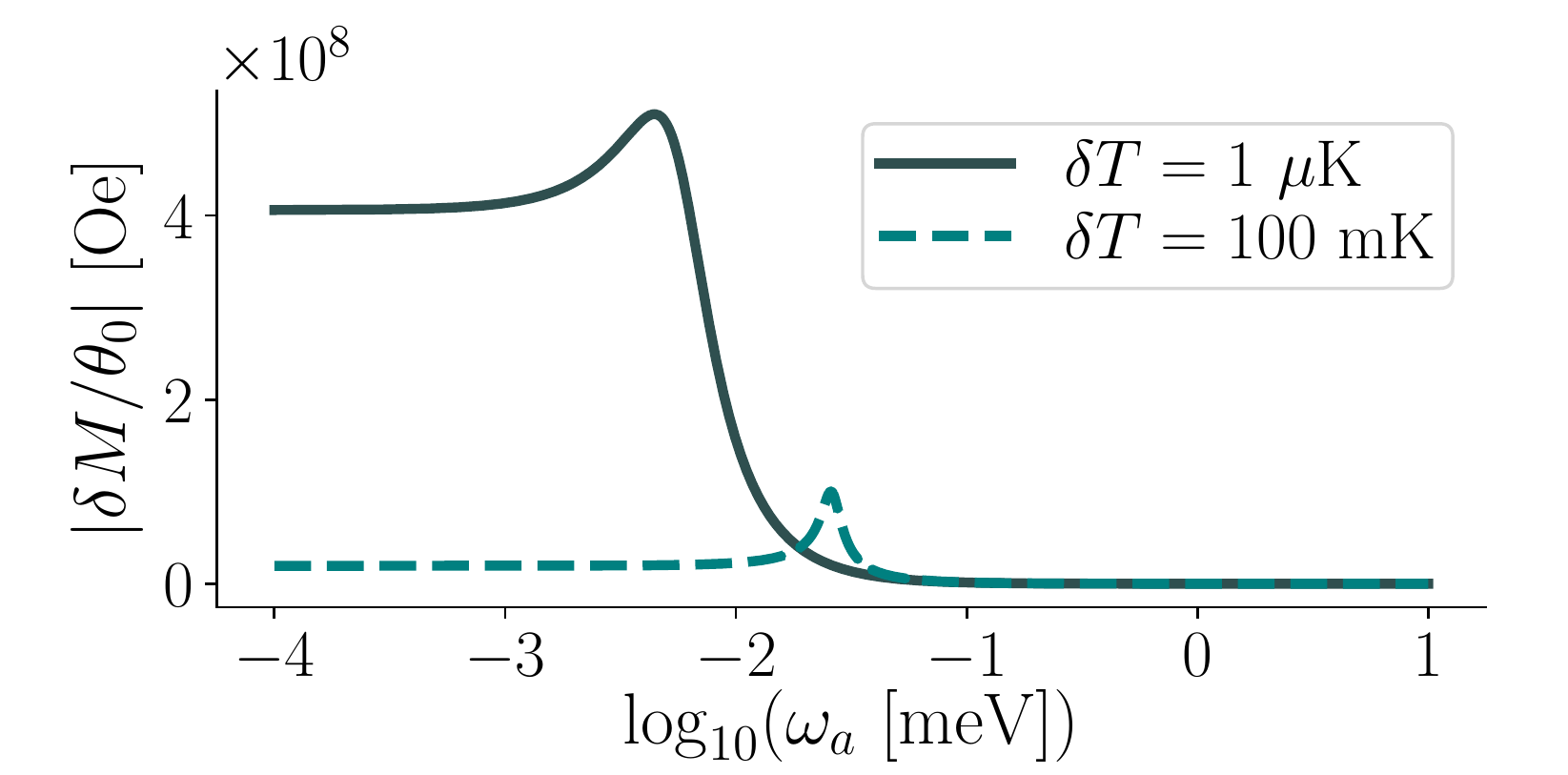}
	\caption{The magnetic linear response for axions coupling to ferroic orders, for two values of $\delta T = T_{C_M} - T$. Dynamical frequency response from Eq.~\eqref{eq:MagresponseMain} assuming that $\alpha_M^{-1}  = (1~\mathrm{meV})^2$ and $\eta_M = 5~\mu$eV. Here we used the parameters $\chi = 1$, $m_P = 4.5\times 10^{-2}$, $\alpha_0 = 10^{-3}$, $b = 0.4$, $\gamma_M = 10^{-5}~\text{K}^{-1}$, $\lambda_M = 9.25~\mathrm{kOe}^{-2}$, $P_0 = 0.7~\mu$C/cm$^2$, and $T_{C_M} = 160~\text{K}$, as roughly motivated by the compound $h$-LSFO, see the main text and Sec.~\ref{sec:Materials}.}
	\label{fig:MagneticSensitivity}
\end{figure}

Secondly, Eq.~\eqref{eq:MagresponseMain} has the characteristic frequency dependence of a driven oscillator. Considering the response as a function of axion frequency, there are three regimes, each of which can be clearly seen in Fig.~\ref{fig:MagneticSensitivity}: 
\begin{enumerate}
    \item $\omega_a \gg \omega_M$:
        The antiadiabatic regime discussed above. The response dies off quickly as the axion frequency becomes much larger than the magnon frequency, so this limit is not optimal. 
    \item $\omega_a \simeq \omega_M$:
        In this regime the response is controlled by the magnon resonance. The response will go as the size of the peak, which will be controlled by the damping coefficient $\eta_M$.
    \item $\omega_a \ll \omega_M$:
        The adiabatic regime. The axion oscillates much more slowly than the magnetic response of the material, therefore it can be approximated as a static external field. Setting $\omega_a=0$ in Eq.~\eqref{eq:MagresponseMain}, the left hand side behaves as $\omega_M^{-2}$. The response will increase as the temperature approaches the critical magnetic temperature because $\omega_M$ becomes small there. 
\end{enumerate}

As discussed in \ref{sec:CritSlowdown}, avoiding the first regime introduces a limitation on what range of axion frequencies can be spanned for a given material and temperature (approximately $\omega_a \lesssim 0.1~$eV).
The axion is expected to be a long-wavelength field, and so the relevant magnon frequency is evaluated at the Brillouin zone centre, $\omega_M=\omega_M(\Gamma)$~\cite{LeinerEA18, SongEA19}.

In the second regime, damping will be a critical factor. The third regime is the one most likely to be accessible in experiment, and the one on which the authors would like to concentrate. It has an attractive feature: The temperature can be used as a tuning dial to adjust both the range of accessible axion frequencies and the size of the response. 

In Fig.~\ref{fig:MagneticSensitivity} the frequency response Eq.~\eqref{eq:MagresponseMain} is shown for parameters motivated by $h$-LSFO. Both the static response enhancement near criticality and the resonance variation with temperature are clearly visible. 

Focusing on the static regime, we would like to study the behaviour of the magnetic response near the critical temperature. As already mentioned, the static response is controlled by $\omega_M^{-2}$, which we must therefore write in terms of the Ginzburg--Landau parameters in Eqs.~\eqref{eq:ELeqM} and \eqref{eq:ELeqP}. To proceed we need the full non-linear solution for the magnetization as presented in Appendix~\ref{sec:Static}---here we summarize the key points. The static magnetization satisfies the cubic equation
\begin{equation}
\begin{aligned}
\gamma_M (T-T_{C_M}) M_0 + 2\lambda_M M_0^3 &+ \frac{c}{2}(\alpha+\chi \theta) P_0 \\
&= \frac{1}{2m_P^2} (\alpha + \chi \theta )^2 M_0\qquad . \label{eq:EigensystemFerro}
\end{aligned}
\end{equation}
The solution can generically be expanded to linear order in $\theta$ as
\begin{equation}
M_0(\theta) = M_0(0) + f(T) \chi \theta + \pazocal{O}(\theta^2),
\label{eq:SolutionExpansion}
\end{equation}
where $f(T)$, which is just the static limit of the linear magnetic response, depends on $\lambda_M$, $m_P$, $\alpha$, $P_0$, and $\gamma_M$. 

There is one extra subtlety: owing to restored time reversal symmetry in the paramagnetic phase, $\alpha = 0$ for $T > T_{C_M}$~\cite{VazEA10, MostovoyEA10}, and so $\alpha$ itself has a temperature dependence near criticality. We model this using $\alpha(T) = \alpha_0 (1-T/T_{C_M})^b$ for some positive exponent $b$. This extra scaling makes the $T \rightarrow T_{C_M}$ limit somewhat subtle, but we find that $f(T)\sim (T_{C_M} - T)^{-2b/3}$. Therefore the magnetic response grows in principal without bound as the critical temperature is approached. In practice, disorder will pose a challenge for temperature tuning, which motivates consideration of a broader range of magnetoelectric materials for dark matter detection. 

\subsection{Avogadro scaling}
\label{sec:AScaling}
In ``quantum sensing'' devices one exploits quantum properties such as entanglement to enhance and measure some physical property~\cite{DegenEA17, ZhuangEA18, XiaEA20, HoEA20}. In the case we consider the ground state of the sensor is a coherent MF state with a wave function given by $\ket{\Psi} = \prod_{\ell} \ket{ \bo{P}_{\ell} \parallel \hat{x}, \bo{M}_{\ell} \parallel \hat{x}} $ in each unit cell $\ell$ (more precisely, each operational unit of the coherent state). The coherent state of the sensor occurs as a result of spontaneous symmetry breaking of both parity and time reversal symmetry without external electric or magnetic fields. This fact alone might provide an added advantage in designing robust DM sensing schemes. 

Within a {\em macroscopic coherent state} scheme each unit cell is functioning as a sensor for DM detection, e.g., as illustrated above. Macroscopic coherence  implies the synchronized response of the material. Each unit cell may have a very low sensitivity due to small coupling and size mismatch between the DM field and the unit-cell fields. In a macroscopically coherent quantum state, however, the unit-cell sensors add coherently and provide an Avogadro scaling enhancement of the signal-to-noise ratio, at least in principle.

In the case of MFs as a putative sensor the magnetic dipole moments, i.e.,~electron spins, are aligned with the electric dipole moment. The ground state of the MF is described by a coherent product state with a finite magnetic moment and polarization in each unit cell. The scalar contributions of Eq.~\eqref{eq:AxionElectronLowE} add constructively over a macroscopic volume ideally limited by the length scale $\min\lbrace \lambda_{a}, L_{\text{domain}} \rbrace$, where $L_{\text{domain}}$ is the linear ferroic domain size and $\lambda_a$ is the axion de Broglie wavelength. Since this wavelength is on the order of meters or more~\cite{IrastorzaEA12}, we anticipate that coupling to a macroscopic $\bo{P} \parallel \bo{M}$ domain can still be ensured on appreciable sensor length scales in clean systems. The coherent nature of the response is sketched in Fig.~\ref{fig:AtomicChain}. As an order-of-magnitude estimate (see Appendix~\ref{sec:AxionFermionFEv2}) we expect effective coupling to the axion to be
\begin{equation}
   \chi \theta_{ae} \sim 7.7 \times 10^{-11} g_{ae}, \label{eq:EffectiveCouplingv3Main}
\end{equation}
so that the total deposited energy becomes
\begin{equation}
\delta E_{ae}^{\text{tot}} = L_{\text{domain}}^3~ \chi \theta_{ae} \sqrt{ \frac{\mu_0}{\varepsilon_0} }~\bo{P} \cdot \bo{M}, \label{eq:FullenergyMain} 
\end{equation}
\begin{figure}[t!bh]
	\centering
	\includegraphics[width=0.70\linewidth]{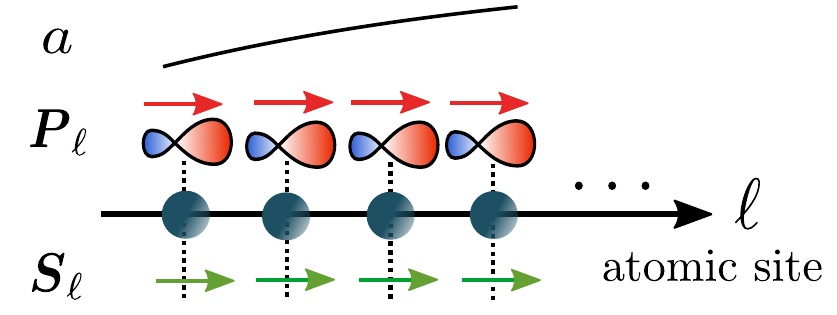} 
	\caption{Atomic sites in a multiferroic where the electron spins $\bo{S}_{\ell}$ are aligned with the electric dipole moments $\bo{P}_{\ell}$. The de Broglie wavelength of the axion $a$ extends to the scale of meters~\cite{IrastorzaEA12}. The coherent (scalar) addition of couplings over atomic sites, given in Eq.~\eqref{eq:AxionElectronLowE}, ideally results in Avogadro scaling of the effective axion coupling.}
	\label{fig:AtomicChain}
\end{figure}
where we have assumed that the axion makes up the local DM density. 

Importantly, though the axion-induced magnetization will be on the order of $\delta M \sim \pazocal{O}(1~\text{aT})$ (in vacuum) due to the small coupling of Eq.~\eqref{eq:EffectiveCouplingv3Main}, the generated magnetic flux scales with the area covered by the magnetometer, $\Phi_{ae} \sim L_{\text{domain}}^2 \delta M$, which can be enhanced experimentally. In practice, the size of the ferroic domains is likely dictated by Kibble--Zurek scaling as we elaborate on in the next Section.

In Appendix~\ref{sec:SNestimate} we provide estimates of the signal-to-noise ratio for SQUID magnetometers, indicating that the Avogadro scaling makes a successful measurement of the response under ideal conditions conceivable. Yet, the tiny magnitude of the proposed effect makes it likely that background noise (e.g.,~thermal or quantum) will overwhelm the signal, and that material growth optimization and signal processing tools will be necessary ingredients if the axion-induced magnetization is to be successfully measured. Finally, we note that in reality decoherence and loss from various sources, such as acoustic phonons and (magnetic) impurities, may make a naive Avogadro scaling difficult to achieve.

\section{Material candidates}
\label{sec:Materials}
Since the discovery of multiferroicity in antiferromagnetic BiFeO$_3$~\cite{WangEA03}, the list of established MFs continues to grow and garner interest for their potential in next-generation microelectronic. However, ferromagnetic order in these materials, which is much sought-after, is rare. Established ferromagnetic cases include CoCr$_2$O$_4$~\cite{YamasakiEA06}, DyFeO$_3$~\cite{TokunagaEA08}, GdFeO$_3$~\cite{TokunagaEA09}, Mn$_2$GeO$_4$~\cite{WhiteEA12, WhiteEA16, HondaEA17, FischerEA20}, and more recently hexagonally grown Lu$_{1-x}$Sc$_x$FeO$_3$ ($h$-LSFO) with $x \approx 0.4$~\cite{LinEA16, DuEA18}. Among these, the  subset with $\bo{P} \parallel \bo{M}$ is practically limited to the latter four, though the polarizations  are not spontaneous in DyFeO$_3$. In Sec.~\ref{sec:hLSFO} we focus on $h$-LSFO by providing an overview of its properties and \emph{ab initio} density functional theory calculations. Then, in Sec.~\ref{sec:OtherCandidates}, we move on to discuss a broader range of other related material platforms, including nonspontaneously symmetry breaking cases.

\subsection{Lu-Sc hexaferrite}
\label{sec:hLSFO}
\begin{figure*}[t!bh]
	\centering
	\includegraphics[width=0.65\linewidth]{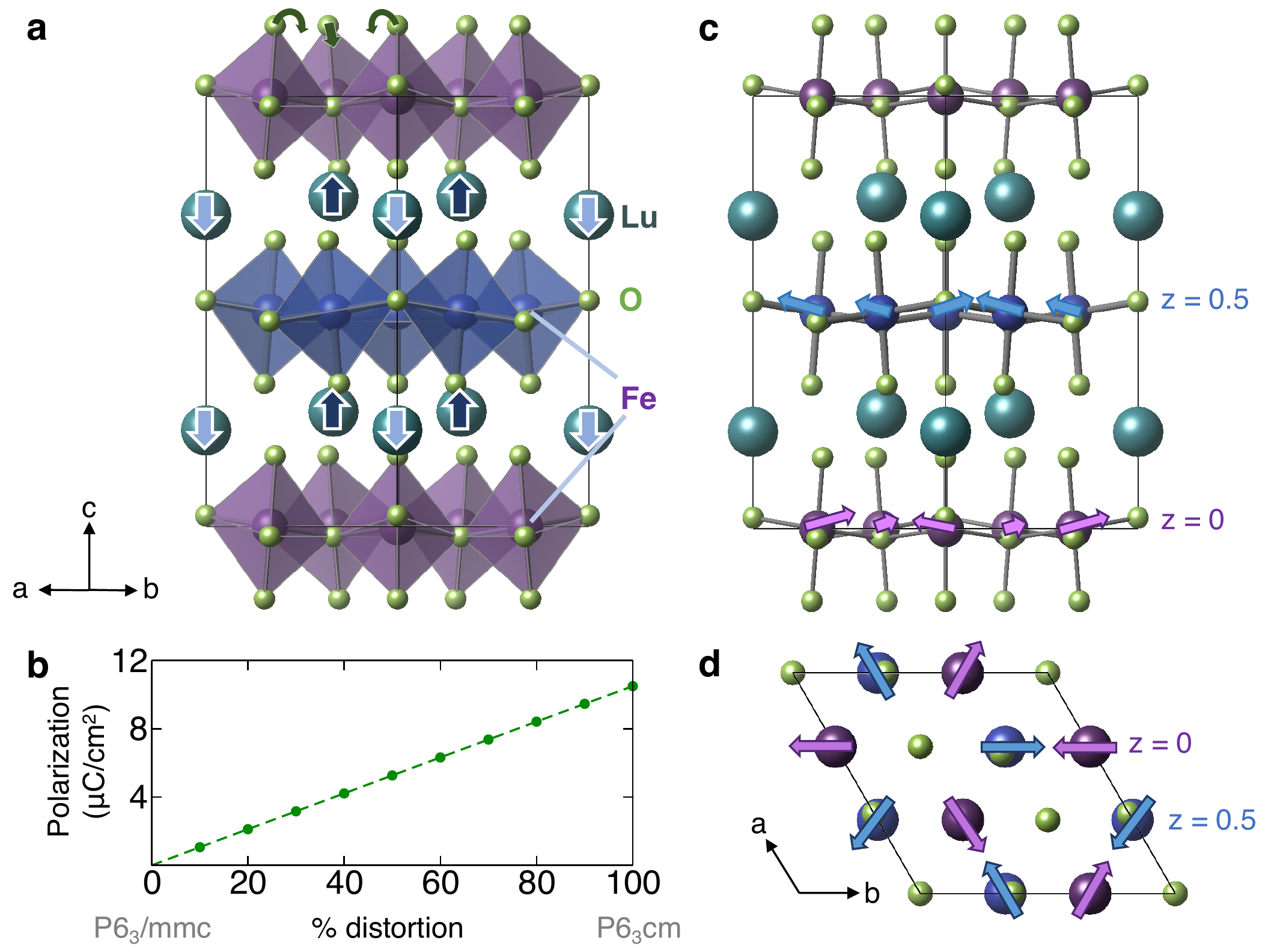}
	\caption{(a) $h$-LuFeO$_{3}$ adopting the polar $P6_{3}$cm space group with the two primary structural changes causing polarization marked---a trimerization and tilting of the Fe-O polyhedra, and a corrugation of the Lu ions---resulting in inversion symmetry breaking. (b) Calculated change in total polarization of $h$-LuFeO$_{3}$ along a path connecting the paraelectric $P6_{3}/mmc$ structure to the ferroelectric $P6_{3}$cm phase. (c), (d) Magnetic order with $A_{2}$ symmetry showing a net out-of-plane magnetization in (c), and compensated frustrated triangular antiferromagnetism in (d).}
	\label{fig:DFT_main}
\end{figure*}

The high-temperature paraelectric phase of the hexagonal ferrite $h$-LSFO adopts the $P6_{3}/mmc$ space group, which has inversion symmetry. At the ferroelectric Curie temperature of $T_{C_P} \approx 1200~\text{K}$, a spontaneous polarization emerges that breaks inversion symmetry. This symmetry-breaking structural distortion is caused by the condensing of a $K_{3}$ phonon mode manifesting in a trimerization of the Fe-O polyhedra, and a polar $\Gamma_{2}^-$ phonon mode causing a rigid shift in the Lu ions. The resulting broken-inversion-symmetry crystal structure adopts the $P6_{3}cm$ space group, as shown in Fig.~\ref{fig:DFT_main} (a), in which the arrows indicate atomic displacement in the ferroelectric phase~\cite{ChoiEA10, DasEA14, DuEA18}. 

In contrast with conventional ferroelectrics, the hexagonal manganites and ferrites form topological defects (vortices) at their ferroelectric phase transition owing to the emergent $U(1)$ symmetry of the structural trimerization at the onset of the ferroelectric phase transition~\cite{Sergey14,Griffin12}. This $U(1)$ symmetry further breaks into a $Z_{6}$ ground state corresponding to the three trimerization directions of both the ``up'' and ``down'' polarization directions, the latter of which can be directly imaged using piezoresponse force microscopy~\cite{Chae12}. Interestingly for our case, the domain sizes are determined by the Kibble--Zurek description of nonequilibrium dynamics during a driven phase transition: In this case the typical standard domain size of $10~\mu\text{m}$ can be controlled by the quench rate through the ferroelectric phase transition~\cite{Griffin12, Griffin17}.

The spin structure of the trimerized Fe$^{3+}$ ions allows for a net magnetization in the $\bo{c}$ direction (in the so-called $A_2$ configuration) due to spin canting as illustrated with purple and blue arrows in Fig.~\ref{fig:DFT_main}(c). Ferromagnetic order develops below $T_{C_M} \approx 160~\text{K}$, with much larger domains of typical size $100~\mu\text{m}$~\cite{DuEA18}. 

To give materials-specific estimates of the expected axion coupling to the matter $\bo{P}$ and $\bo{M}$ we use \textit{ab initio} calculations based on density functional theory (DFT) in the case of hexagonal LuFeO$_{3}$. The calculation details are given in Appendix~\ref{sec:DFTapp}. Here we summarize the key parameters used in the estimates of the strength of the effective matter coupling, namely the expectation value of the parallel polarization $\bo{P}$ and net ferromagnetism $\bo{M}$. 

We first confirmed that the $A_{2}$ magnetic ordering adopted by the $P6_{3}cm$ polar structure is a frustrated 120\degree\ antiferromagnetic ordering with a slight out-of-plane canting. The calculated out-of-plane spontaneous magnetization was found to be 0.02 $\mu_{B}$/Fe, consistent with previous theoretical and experimental reports~\cite{DasEA14, MoyerEA14, DisselerEA15}. We find this value is relatively insensitive to choice of $U_{\mathrm{eff}}$ in our DFT+U calculations for modest values of $U_{\mathrm{eff}}$. However, we note that an alternative magnetic order with $A_{1}$ symmetry is essentially degenerate with the $A_2$ magnetic order with a calculated energy of 0.06 meV/f.u. lower in energy than $A_{2}$, which is also consistent with other experiments~\cite{LiuEA19, LeinerEA18}. Such a small energy barrier between these two magnetic orders and the contradictory experimental measurements point to highly tunable magnetic orders that can be sensitive to synthesis conditions and strain, for example. We briefly note that this opens a possibility for increasing the observed out-of-plane net spontaneous magnetization through strain, chemical substitution, or doping.

Next we calculated the magnitude of the spontaneous polarization of the $P6_{3}cm$ structure using the Berry phase approach. Debate over the definition of macroscopic polarization in bulk periodic systems by summation over local electric dipoles included whether it was truly a bulk effect, or a result of surface termination details, and how to resolve the multivaluedness of the polarization resulting from unit cell choices~\cite{Martin74, Spaldin12}. These issues were resolved with the introduction of the modern theory of polarization, which defined the bulk polarization from the Berry phase of the constituent wave functions, naturally prescribing the gauge freedom arising from the Berry phase~\cite{Resta92, KSmithEA93}. To resolve this latter point in calculations, we generate a set of interpolated structures between the paraelectric $P6_{3}/mmc$ and the ferroelectric $P6_{3}cm$ to give a smooth interpolation of the Berry phase calculated electronic polarization. Our final calculated polarization was $10.5~\mu \mathrm{C}/\mathrm{cm}^2$ in the out-of-plane direction.

\subsection{Other candidates and nonspontaneous symmetry breaking}
\label{sec:OtherCandidates}
As a second promising material platform we mention the olivine compound Mn$_2$GeO$_4$~\cite{WhiteEA12, FischerEA20}. In this material the ferroic orders originate from spiral spin-order, resulting from transverse conical chains that yield a net magnetization and electric polarization pointing along the $\bo{c}$ direction. Moreover, magnetic order here sets in at $47~\text{K}$. Lowering the temperature leads to two additional first-order transitions, and below $T_{C_P} \approx 5.5~\text{K}$ both $\bo{M} \parallel \bo{P} \parallel \bo{c}$ are spontaneously generated. At low temperature the magnitude of the spontaneous electric and magnetic polarizations along $\bo{c}$ are about $6~\mu$C$/$m$^2$ and $7\times 10^{-3}~\mu_B/\text{Mn}$, respectively~\cite{WhiteEA12}. Ferroelectric domains on the scale of $500~\mu\text{m}$ have been observed in this compound~\cite{LeoEA18}. 

Another possibility is to relax the assumption of spontaneous polarizations and consider material candidates where the magnetic $A_2$ state is induced by an external magnetic field, though we note that such cases lie beyond the scope of the theory in Eq.~\eqref{eq:FreeEnergy}. The realization of a field-induced $A_2$ phase was proposed for the hexagonal manganites $h$-$R$MnO$_3$ with $R =\text{Ho},~\text{Er},~\text{Tm},~\text{Yb}$~\cite{FiebigEA03}. For instance, $h$-ErMnO$_3$ is believed to realize a field-induced ferromagnetic phase for a broad range of external magnetic fields. Furthermore, this may allow the reach of favourably large polarizations, of several $\mu_B/\text{Er}$ and $\mu\text{C}/\text{cm}^2$. However, the details of the magnetic phase diagram are complicated and still debated~\cite{MeierEA12, LiuEA18}.

Antiferromagnetic materials naively appear unsuited for sensing the axion-matter coupling [of Eq.~\eqref{eq:AxionElectronLowE}] since the axion would couple with alternating sign to the anti-aligned spins and average out. However, ME field cooling techniques have demonstrated controlled switching of the antiferromagnetic state in Cr$_2$O$_3$ heterostructures and made single ME domains possible in principle~\cite{BorisovEA05, XiEA10}. Hence, even antiferromagnetic magnetoelectrics may be surveyed for axion detection~\cite{KakhidzeEA10}.

Finally, we mention the possibility of considering heterostructures with alternating layers of ferromagnetically and ferroelectrically ordered constituents, which also could enhance the space of relevant materials platforms.

%
%%
%%%
\section{Conclusion and outlook}
\label{sec:Conc}
%%%
%%
%

In this paper we have considered coupling between axion dark matter and macroscopic orders in multiferroics. The proposed scheme fits into the growing set of proposals to use quantum materials where the (broken) symmetry of the ground state enables the coupling between the axion and matter fields. Key advantages of the proposed MERMAID platform for DM detection include (i) reliance on the spontaneously broken symmetry state rather than external fields, and (ii) Avogadro scaling that allows long-range crystalline order in nature to build a coherent stack of sensors instead of experimenters. 

Using materials-specific quantities, we estimated the axion-electron coupling in multiferroic materials with parallel magnetic and electric polarizations. The effective axion coupling in these materials is expected to be of the form $\bo{P} \cdot \bo{M}$, similar to a linear magnetoelectric effect. We argued that the coupling is enhanced/resonant as the longitudinal magnon frequency matches the axion frequency. While we use an established MF material for these estimates, further studies should explore how to enhance the linear polarizations, and hence the coupling to axions, in real materials. Moreover, while here we focus on the case of spontaneously broken symmetries to induce electric- and magnetic-dipoles without using any external magnetic or electric fields, one can enhance the magnetization and polarization in the material with applied fields. 

Temperature may in principle be used as a tuning knob to scan axion frequencies and to boost the response near critically. The sensing is based on extreme sensitivity of the multiferroic magnetic response near the ferromagnetic transition to the parameter changes induced by the axion field. In this respect the idea is similar to the transition-edge sensor.

We discussed Avogadro scaling as a result of the coherent response in materials with macroscopic quantum order. The quantum coherence of all unit cells means that the amplitude for scattering at each cell in the coherent state effectively enhances the signal-to-noise ratio. In the case of multiferroics this coherence emerges as a result of a macroscopic ferromagnetic and ferroelectric state.  

High sensitivity magnetometers such as SQUIDs combined with discoveries of $\bo{P} \parallel \bo{M}$ multiferroics, such as the ferrites Lu$_{1-x}$Sc$_x$FeO$_3$ and the olivine compound Mn$_2$GeO$_4$, may provide a realization of the axion-matter coupling. In future work we would like to expand the sensitivity analysis and address the expected limitations in greater detail. As identification of suited condensed matter systems for axion and dark matter detection is in its infancy, exploitation of multiferroics for this purpose has the potential of opening a new avenue of research. There is also a possibility of considering dark matter detection schemes via the axion-matter coupling in a wider class of magnetoelectric materials than in the fairly limited class considered here. \newline

\begin{acknowledgments}
We thank Frank Wilczek, Alexander John Millar, Matthew Lawson, Yonathan Kahn, Ilya Sochnikov, Stefano Bonetti, Jan Conrad, Alfredo Ferella, Andrew Geraci, Johan Hellsvik, Rohit Prasankumar, Konstantin Beyer, Felix Flicker, and Nicola Spaldin for useful discussions. 

The work was supported by the research environment grant ``Detecting Axion Dark Matter In The Sky And In The Lab (AxionDM)'' funded by the Swedish Research Council (VR) under Dnr 2019-02337. H.S.R., B.F., S.B., A.M., and A.V.B.~were also supported by the University of Connecticut, the European Research Council under the European Unions Seventh Framework ERS-2018-SYG 810451 HERO and VILLUM  FONDEN  via  the  Centre  of  Excellence  for  Dirac Materials (Grant No.~11744). S.M.G.~was supported by the Quantum Information Science Enabled Discovery (QuantISED) for High Energy Physics (KA2401032). This work used the Extreme Science and Engineering Discovery Environment (XSEDE), which is supported by National Science Foundation Grant No.~ACI-1548562. Work at the Molecular Foundry was supported by the Office of Science, Office of Basic Energy Sciences, of the U.S. Department of Energy under Contract No. DE-AC02-05CH11231. S.-W.C.~was supported by the DOE under Grant No. DOE: DE-FG02-07ER46382. 
\end{acknowledgments}
\bibliography{MultiferroicsDM}

\newpage

%
%%
%%%
%%%%
%\newpage
%\counterwithin{figure}{section}

\begin{appendix}
\onecolumngrid

%
%%
%%%
\section{The axion-fermion coupling}
\label{sec:AxionFermion}
%%%
%%
%
The axion couples to fermions via (with $c = \hbar = 1$)~\cite{GondoloEA09, BarthEA13, Sikivie14}
\begin{equation}
\pazocal{L} = \bar{\psi}\left( i \gamma^{\mu} D_{\mu} - m_f \right) \psi - g_{af}  \frac{\partial_{\mu} a}{2m_f} \bar{\psi} \gamma^5 \gamma^{\mu} \psi,
\label{eq:LagAxionFermion}
\end{equation}
where $D_{\mu} = \partial_{\mu} - i q A_{\mu}$ is the covariant derivative, $g_{af}$ is a dimensionless coupling strength, and $q$ and $m_f$ are the fermion charge and mass, respectively. Moreover, $\gamma^{\mu}$ are Dirac matrices in a  particle-hole spinor basis $\psi = \left( \psi_f, \psi_{\bar{f}} \right)$. The Lagrangian of Eq.~\eqref{eq:LagAxionFermion} yields the equations of motion
\begin{align}
\left[ E + q\varphi - m_f  + g \bo{\nabla} a \cdot \bo{\sigma} \right] \psi_f &= \left[ - g \partial_t a + \bo{\sigma} \cdot (\bo{p} - q \bo{A}) \right] \psi_{\bar{f}}, \label{eq:FermionEq} \\
\left[ E + q\varphi + m_f + g \bo{\nabla}a \cdot \bo{\sigma} \right] \psi_{\bar{f}} &= \left[ - g \partial_t a + \bo{\sigma} \cdot (\bo{p} - q \bo{A}) \right] \psi_{f}, \label{eq:AntiFermionEq}
\end{align}
where $g \equiv g_{af} / (2m_f)$, $A = (\varphi, \bo{A})$, and $i \partial_0 \psi_f = E \psi_f$, and $\bo{\sigma}$ is a vector of Pauli matrices. In the nonrelativistic limit, with $E \approx m_f$, and $q\varphi \ll m_f$, these equations give the effective low-energy Hamiltonian (cf.~Ref.~\cite{Sikivie14})
\begin{equation}
H_{af} = \frac{1}{2m_f} \left[ \bo{\sigma} \cdot (\bo{p}-q\bo{A}) \right]^2 - \frac{g_{af}}{2m_f}\left[ \bo{\nabla} a \cdot \bo{\sigma} + \frac{\partial_t a}{m_f} \bo{\sigma} \cdot (\bo{p}-q\bo{A})  \right] + \pazocal{O}(g_{af}^2, \partial^2 a ),
\label{eq:HamEffective}
\end{equation}
where the left out terms are either higher order in $g_{af}$ or in derivatives of $a$ or in both. The $\partial_t a$ term is expected to be suppressed by $m_a/(2m_f)$ (which is $\lesssim 10^{-7}$ for electrons)---nevertheless this is the term, which we choose to focus on.  The so-called ``axion wind'' coupling $\sim \bo{\nabla} a$ term may be relevant some other settings~\cite{Sikivie14, Hill16, MitridateEA20}. 

%
%%
%%%
\section{Axion-electron coupling in multiferroics}
\label{sec:AxionFermionFEv2}
%%%
%%
%
Here we consider the $\partial_t a$ term of the coupling in Eq.~\eqref{eq:HamEffective}:
\begin{equation}
\begin{aligned}
H_{ae} &= - i a_0 \tilde{g}~\bo{p}_j \cdot \bo{\sigma}_j ,  \\
\tilde{g} &\equiv g_{ae} \frac{m_a}{2m_e^2 c},
\end{aligned}
\label{eq:AxionElectronEffective}
\end{equation}
where the factor of $m_a$ appears when assuming an axion field $a = a_0 \exp(i m_a t)$, which is valid since the de Broglie wavelength is on the order of meters~\cite{IrastorzaEA12}. As explained in Sec.~\ref{sec:Axions} the magnitude of $a_0$ is fixed by assuming that the axion make up the local DM density~\cite{MillarEA17, CatenaUllio10, PDG18}, $\rho_{\text{DM}} = \frac{1}{2} m_a^2 \lvert a_0 \rvert^2 \approx 300~\text{MeV} \text{cm}^{-3}$. 

\begin{figure}[h!tb]
	\centering
	\includegraphics[width=0.35\linewidth]{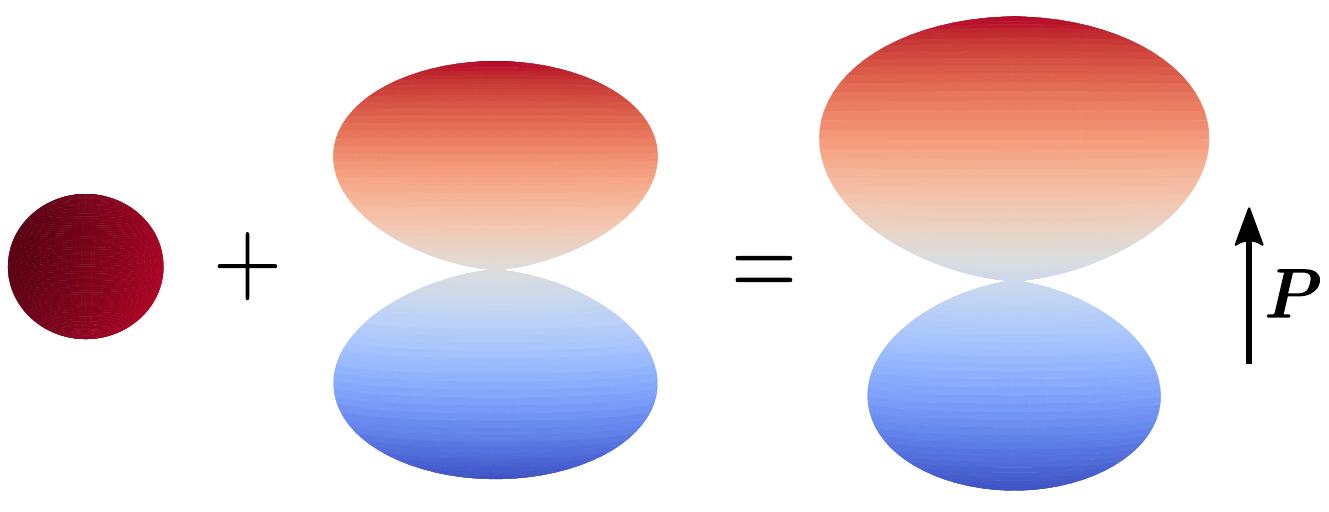} 
	\caption{Illustration of the ferroelectric polarization $\bo{P}$ generated by a mixed orbital state, consisting of $\Psi_+$ ($s$-wave) and $\Psi_-$ ($p_z$-wave), with color representing charge density.}
	\label{fig:Orbitals}
\end{figure}

To quantify the size of the effect induced by Eq.~\eqref{eq:AxionElectronEffective} one should ideally calculate
\begin{equation}
\delta E_{ae} = - i a_0 \tilde{g} \bra{\Psi}\bo{\sigma} \cdot \bo{\nabla} \ket{\Psi}, \label{eq:ExpectationAE} 
\end{equation}
where $\ket{\Psi}$ is the many-body electronic ground state [i.e.,~$\Psi = \pazocal{A} \Psi({\bf r}_1, s_1; {\bf r}_2, s_2;\dots;{\bf r}_N, s_N)$, $\pazocal{A}$ being the antisymmetrizing operator] with spin-orbit coupling included, which is a nontrivial problem. For the material estimates in Sec.~\ref{sec:Materials} and Appendix~\ref{sec:DFTapp} $\ket{\Psi}$ is determined with spin-orbit coupling included self-consistently using density functional theory. Here we take a simplified approach to give an order-of-magnitude estimate for the nonrelativistic outer electron orbitals. For the expectation value of Eq.~\eqref{eq:AxionElectronEffective} one can generally insert a resolution of the identity such that
\begin{equation}
\delta E_{ae}  = - i a_0 \tilde{g} \Big[ \bra{\Psi} \bo{p}  \ket{\Psi} \cdot \bra{\Psi} \bo{\sigma} \ket{\Psi} + \sum_{n \neq \Psi}  \bra{\Psi} \bo{p}  \ket{n} \cdot \bra{n} \bo{\sigma} \ket{\Psi} \Big].
\label{eq:AxionElectronNew1}
\end{equation}
In the subsequent steps we ignore the off-diagonal terms in Eq.~\eqref{eq:AxionElectronNew1}. For an insulating state the excited states are separated by the gap and their contributions to Eq.~\eqref{eq:AxionElectronNew1} are expected to be small. A realistic estimate taking them into account will have to be done using \emph{ab initio} methods and will be a subject of subsequent work. We thus approximate
\begin{equation}
\delta E_{ae}  \approx -  i a_0 \tilde{g} \bra{\Psi} \bo{p} \ket{\Psi} \cdot \bra{\Psi} \bo{\sigma} \ket{\Psi}.
\label{eq:AxionElectronEffectiveNew}
\end{equation}
To have a ferroelectric polarization, i.e.,~to account for the misalignment of charge centres, one needs a linear combination of an even and odd orbital parity component, labelled by subscript $\pm$:
\begin{equation}
\ket{\Psi} = a \ket{\Psi_+} + b \ket{\Psi_-},
\label{eq:LinearCombination}
\end{equation}
as illustrated in Fig.~\ref{fig:Orbitals}, resulting in 
\begin{equation}
\bra{\Psi} \bo{p} \ket{\Psi} = -2i\hbar \Re{ a^{\ast} b } \bra{\Psi_+} \bo{\nabla} \ket{\Psi_-}.
\label{eq:Overlap}
\end{equation}

The electric polarization takes the form $\bo{P} = \rho_e \int \D\bo{r}~ \bo{r} \lvert \Psi(\bo{r})\rvert^2 =  \rho_e \bra{\Psi} \bo{r} \ket{\Psi}$, where $ \rho_e$ is the charge density. This noticeably vanishes for any pure orbital state. However, for the mixed orbital state of Eq.~\eqref{eq:LinearCombination} we get
\begin{equation}
    \bo{P} = 2 \rho_e \Re{ a^{\ast} b } \bra{\Psi_+} \bo{r} \ket{\Psi_-} \propto \frac{e}{a_B} \bra{\Psi} \bo{p} \ket{\Psi},
\label{eq:PolarizationMixed}
\end{equation}
if the (atomic) spatial variations take place over the scale of the Bohr radius, $a_B$. Hence Eq.~\eqref{eq:PolarizationMixed} shows that essentially 
\begin{equation}
\bo{P} \simeq \frac{2e}{a_B} \Re{ a^{\ast} b } \bra{\Psi_+} \bo{\nabla} \ket{\Psi_-}. 
\label{eq:EffectivePolarization}
\end{equation}
The spin expectation value defines a magnetic moment in the ferromagnetic phase, $\bra{\Psi} \bo{\sigma} \ket{\Psi} = -\frac{2}{g_s \mu_B} \bo{\mu}$, where $g_s$ is the effective electron $g$-factor, $\mu_B$ is the Bohr magneton, and $\bo{\mu}$ is the total magnetic moment of the unit cell. The microscopic contributions are summed over individual spins in the unit cell (giving $\bo{\mu}$). In the ferromagnetic state we get
\begin{equation}
 \delta E_{ae} \approx a_0 \tilde{g} \frac{2 \hbar}{g_s \alpha e^2} \bo{P} \cdot \bo{\mu} = a_0 \tilde{g} \frac{2 \hbar}{g_s \alpha e^2} V_{\text{uc}}~\bo{P} \cdot \bo{M},
 \label{eq:CouplingFinal}
\end{equation}
where $\alpha$ is the fine structure constant and the subscript ``uc'' refers to the unit cell. This demonstrates how a macroscopic $\bo{P} \cdot \bo{M}$ coupling results from the axion-fermion coupling in a multiferroic. The above contributions add up coherently over a macroscopic number of unit cells, as dictated by the ferroic domain size. When comparing to the definition of $\chi \theta_{ae}$ in Eq.~\eqref{eq:FreeEnergy}, we arrive at the effective dimensionless coupling:
\begin{align}
\chi \theta_{ae} &\equiv g_{ae}  \frac{a_0 m_a \hbar}{ m_e^2 c^2 g_s \alpha e^2} \sqrt{ \frac{\varepsilon_0}{\mu_0} } \sim 7.7 \times 10^{-11} g_{ae}, \label{eq:EffectiveCouplingv3} \\
\delta E_{ae}^{\text{tot}} &\approx L_{\text{domain}}^3~ \chi \theta_{ae} \sqrt{ \frac{\mu_0}{\varepsilon_0} }~\bo{P} \cdot \bo{M}. \label{eq:Fullenergy}
\end{align}
Though the effective coupling is small, the total energy scales with volume and hence takes advantage of the coherent state of the material. For representative parameters $g_{ae} = 10^{-13}$, $M_0 = 10~$Oe, $P_0 = 0.7$~$\mu$C/cm$^2$, and $L_{\text{domain}} = 1~$mm the deposited energy from the axion is $\delta E_{ae} = 0.1~$neV.

%
%%
%%%
\section{Relativistic corrections}
\label{sec:rel}
%%%
%%
%
The first relativistic correction to the Hamiltonian of Eq.~\eqref{eq:AxionElectronLowE} in natural units is
\begin{equation}
    H_{\text{rel}}\, =\, -\frac{i}{4}\frac{g_{ae}}{m_e^2}(\partial_t a)\frac{\bo{\nabla}^2}{m_e^2} (\bo{\sigma} \cdot \bo{\nabla}).
    \label{eq:secondcorrection}
\end{equation}
The scale of this correction with respect to the leading term we consider,  Eq.~\eqref{eq:AxionElectronEffective}, is on the order of 
\begin{eqnarray}
   \langle H_{\text{rel}}/H_{ae} \rangle \sim \frac{Z^2 e^2}{a_B m_e c^2} \sim 10^{-1},
   \label{eq:relativestrength}
\end{eqnarray}
which holds for the deepest electronic states of Lu ($ Z = 71$), and with $a_B$ being the Bohr radius, and $H_{ae}$ given by Eq.~\eqref{eq:AxionElectronEffective}. Hence we conclude that for materials of interest the next-to-leading order term in the relativistic expansion of the axion-electron coupling is subdominant.

%
%%
%%%
\section{Details of the dynamics}
\label{sec:Mathieu}
%%%
%%
%
We consider spatially homogeneous solutions of the longitudinal modes of Eqs.~\eqref{eq:ELeqM} and \eqref{eq:ELeqP}, and ignore the time dependence of $\mathbf{P}$:
\begin{equation}
\begin{aligned}
\mathbf{P}(t,\mathbf{x})& = P\, \hat{e}, \\
\mathbf{M}(t,\mathbf{x}) &= M(t)\, \hat{e}, 
\end{aligned}
\end{equation}
with $\hat{e}$ being a unit vector dictating the direction of the spontaneous magnetization/polarization. The dynamics is governed by the coupled equations
\begin{equation}
\begin{aligned}
\alpha_M \ddot M +\gamma_M(T-T_{C_M}) M + 2 \lambda_M M^3 + \frac{c}{2} (\alpha+\chi\theta)P\, &= \, 0, \\
\gamma_P(T-T_{C_P}) P + 2 \lambda_P P^3 + \frac{1}{2c} (\alpha+\chi\theta)M\, &= \, 0.
\label{eq:DiffOp}
\end{aligned}
\end{equation}
Denote by $M_0$ and $P_0$ the values of the static polarizations in the absence of the axion field. These solve equations \eqref{eq:DiffOp} with the time derivative and $\theta$ set to zero. Now linearizing Eqs.~\eqref{eq:DiffOp} around this static solution, $M=M_0+\delta M$, $P=P_0+\delta P$ produces
\begin{align}
\alpha_M \ddot{\delta M} + m_M^2 \delta M + \frac{c}{2}[(\alpha+\chi\theta)\delta P + P_0 \chi\theta] \, &= \, 0, \\
m_P^2\delta P + \frac{1}{2c}[(\alpha+\chi\theta)\delta M + M_0 \chi\theta] \, &= \, 0,
\end{align}
where we defined the (dimensionless) effective masses about equilibrium:
\begin{align}
    m_M^2 &\equiv  \gamma_M(T-T_{C_M}) + 6 \lambda_M M_0^2, \\
    m_P^2 &\equiv \gamma_P(T-T_{C_P}) + 6 \lambda_P P_0^2 .
\end{align}
Now we solve for $\delta P$, and also neglect terms $\propto \delta P\, \theta, \delta M\, \theta$ as second order quantities. Thus we arrive at an equation in the form of a standard driven harmonic oscillator
\begin{align}
    & \ddot{\delta M} + \eta_M \dot{\delta M} + \omega_M^2 \delta M \, = \, \frac{1}{2\alpha_M} \left( \frac{\alpha M_0}{2 m_P^2} - c P_0\right)\chi\theta, \label{eq:sho0} \\
    & \omega_M^2 \equiv \frac{1}{\alpha_M}\left[ m_M^2 - \frac{\alpha^2}{4m_P^2} \right].
    \label{eq:sho1}
\end{align}
We have inserted by hand a phenomenological damping term $\sim \eta_M$ into Eq.~\eqref{eq:sho0}, which does not follow from the preceding equations but is expected on physical grounds, to model dissipation in the system. The driving term is proportional to the axion field. This is expected to oscillate, $\theta = \theta_0 \exp (i \omega_a t)$, with the frequency associated to its rest mass energy $\omega_a = (c^2/\hbar) m_a$. We may therefore write down the linear response of the magnetization, as a function of the frequency: 
\begin{equation}
    \Big\lvert \frac{\delta M}{\theta_0}\Big \rvert \, =\, \frac{1}{\sqrt{ \left(\omega_a^2 - \omega_M^2\right)^2+ \eta_M^2 \omega_a^2 }}\, \frac{\chi}{2\alpha_M} \Big\lvert \frac{\alpha M_0}{2 m_P^2} - c P_0 \Big\rvert.
    \label{eq:magresponse}
\end{equation}
The first term on the right hand side is the standard factor displaying the resonance between the magnetic response and the axion field when $\omega_a = \omega_M$. A key question is how close we can tune the properties of our system to maximize the response function of Eq.~\eqref{eq:magresponse}.

%
%%
%%%
\section{Static solution}
\label{sec:Static}
%%%
%%
%
The full expressions for the first terms in the series expansion of $M_0(\theta)$ as appearing in Eq.~\eqref{eq:SolutionExpansion} are given by
\begin{equation}
\begin{aligned}
M_0(\theta) &= M_0(0) + \chi \theta f + \pazocal{O}(\theta^2), \\
M_0(0) &= \frac{3^{2/3} \alpha ^2 \lambda_M +\sqrt[3]{3} m_P^2 \left(\left(l-9 \alpha  c \lambda_M^2 P_0 \right)^{2/3}+2 \sqrt[3]{3} \gamma_M
   \lambda_M  (T_{C_M}-T)\right)}{6 \lambda_M  m_P^2 \left(l-9 \alpha  c \lambda_M^2 P_0\right)^{1/3}}, \\
   f &= \Bigg[ \frac{\alpha  l \left(9 c^2 P_0^2 m_P^6 \lambda _M-\left(\alpha ^2+2 m_P^2 \gamma _M
   \left(T_{C_M}-T\right)\right)^2\right)}{\lambda _M \left(27 \alpha ^2 c^2 P_0^2 m_P^6 \lambda _M-\left(\alpha ^2+2 m_P^2
   \gamma _M \left(T_{C_M}-T\right)\right)^3\right)} \\
   &\hspace{20pt}-\frac{2 \sqrt[3]{3} \left(\frac{\alpha ^2}{2 m_P^2}+\gamma _M
   \left(T_{C_M}-T\right)\right) \left(\frac{\alpha  l \left(9 c^2 P_0^2 m_P^6 \lambda _M-\left(\alpha ^2+2 m_P^2 \gamma _M
   \left(T_{C_M}-T\right)\right)^2\right)}{27 \alpha ^2 c^2 P_0^2 m_P^6 \lambda _M-\left(\alpha ^2+2 m_P^2 \gamma _M
   \left(T_{C_M}-T\right)\right)^3}-3 c P_0 \lambda _M^2\right)}{\left(l-9 \alpha  c P_0 \lambda _M^2\right)^{2/3}} \\
   &\hspace{20pt}+\frac{2
   \sqrt[3]{3} \alpha  \sqrt[3]{l-9 \alpha  c P_0 \lambda _M^2}}{m_P^2}-3 c P_0 \lambda _M \Bigg] \Big/ \Big[  2\ 3^{2/3} \left(l-9 \alpha  c P_0
   \lambda _M^2\right)^{2/3} \Big], \\
   l &= \frac{1}{m_P^3}\left[ 81 \alpha ^2 c^2 P_0^2 m_M^6 \lambda _M^4-3 \lambda _M^3 \left(\alpha ^2+2 m_P^2 \gamma _M
   \left(T_{C_M}-T\right)\right)^3\right]^{1/2}.
\end{aligned}
\label{eq:StaticExplicit}
\end{equation}

\section{Signal-to-noise ratio}
\label{sec:SNestimate}
To give an estimate of the signal-to-noise ratio for the schematic setup of Fig.~\ref{fig:Idea}, we here compare the magnitude of the axion induced magnetic field to presently achievable sensitivity levels for white noise in SQUIDs.

From the response of Eq.~\eqref{eq:MagresponseMain}, we obtain a (vacuum) magnetic response of magnitude $\delta M \approx 0.35$~aT ($\delta M \approx 0.31$~aT) at $\omega_a = \omega_M$ ($\omega_a \ll \omega_M$), i.e.,~on (off) resonance, at $T-T_{C_M} = 1~\mu$K and using the parameters of Fig.~\ref{fig:MagneticSensitivity}, which are motivated by the compound $h$-LSFO. We reiterate, however, that at resonance the response is controlled by the loss rate $\delta M \propto 1/\eta_M$, which experimentally may well be the limiting factor. 

A key aspect of our proposal is that our signal will add \textit{coherently} over the surface area of the sample, limited only by the maximum ferromagnetic domain size available. Given a linear domain size of $1~\mathrm{mm}$, we estimate the total magnetic flux measured over the whole area of the surface ($1~\mathrm{mm}^2$). This gives on the order of $10^{-8}\Phi_0$, where $\Phi_0\equiv (hc)/2e$ is the fundamental flux quantum. 

On the other hand, presently available SQUID technologies~\cite{StormEA16, BuchnerEA18, Gramolin20, SochnikovEA20} can reach white noise sensitivities as low as $12~\text{n}\Phi_0 /\sqrt{\text{Hz} }$ (though $1/f$ noise, of which origin is not fully understood, usually limits the sensitivity at frequencies $f \lesssim 1~$ kHz). Therefore we find that in order to reach a signal-to-noise ratio of $S/N\sim 1$, we require a bandwidth of $\sim 1~\mathrm{Hz}$, and so a total measurement integration time of $\sim 1~\mathrm{s}$. 

Both loss and the bandwidth dependence of $S/N$ likely are more restrictive in practice~\cite{BudkerEA14}. Finally, we note that impurities may be another relevant and leading threat against the success of the scheme proposed here. Similar issues may, however, be even more pressing in doped topological insulators, which tend to suffer from residual bulk conductance and charge puddles~\cite{BeidenkopfEA11}.

%
%%
%%%
\section{Density functional theory calculation details}
\label{sec:DFTapp}
%%%
%%
%

Our density functional theory (DFT) calculations were performed with the Vienna \emph{ab initio} simulation package (VASP) with projector augmented waves and the PBE exchange-correlation functional. We treated Lu ($5s$, $5p$, $6s$), Fe ($3d$, $4s$, $4p$), O ($2s$, $2p$) electrons as valence. We used a plane wave energy cutoff of 600 eV and a $\Gamma$-centered $k$-point grid of $6 \times 6 \times 2$. Since semilocal density functionals suffer from the well-documented self-interaction error, we use DFT+U to effectively treat the Fe-$d$ orbitals. In this case, we define an effective $U_{\mathrm{eff}} \equiv U-J$ and select a value of $U_{\mathrm{eff}}$ = 4 eV for the Fe-$d$ orbitals, consistent with previous studies on h-LuFeO$_{3}$~\cite{DasEA14}.

We used the PBEsol functional for structural optimizations, which is specifically tailored for accurate forces in solids~\cite{PBEsol}. Structural optimizations of the unit cell shape and size in addition to the internal coordinates were performed until the Hellman--Feynman forces were less than 0.01 eV/\AA~with the resulting lattice parameters  given in the Table \ref{table:latticepms}. The spontaneous polarization was calculated using the Berry-phase approach with interpolated structures between the nonpolar and polar hexagonal structures used to obtain consistent branch cuts for the calculated polarization. Spin-orbit coupling was included self-consistently for all calculations.
\begin{table}[h]
\centering
\caption{Lattice parameters for the high-symmetry $P6_{3}/mmc$  and low-symmetry $P6_{3}cm$  phases of LuFeO$_{3}$ calculated using PBEsol. The last row are the measured lattice parameters from experiment taken from Magome \emph{et al.}~\cite{MagomeEA10}.}
\label{table:latticepms}
\begin{tabular}{ p{2.5cm}  p{3.0cm} p{3.0cm} }
\toprule
\textbf{Space Group} & \textbf{a (\AA)}                                        & \textbf{c (\AA)}                                          \\ \hline
$P6_{3}/mmc$              & 3.438 (Th.)                                                 & 11.592     (Th.)                                              \\ \hline
$P6_{3}cm$                & \begin{tabular}[c]{@{}l@{}}5.941 (Th.) \\ 5.965 (Exp.) \end{tabular} & \begin{tabular}[c]{@{}l@{}}11.523 (Th.)\\ 11.702  (Exp.) \end{tabular} \\ \hline \hline
\end{tabular}
\end{table}

\end{appendix}

\end{document}